\documentclass[nopreprint%
]{jasatex}
\usepackage{graphicx}
\usepackage{dcolumn}
\usepackage{amsmath,amsfonts}
\usepackage{bm}
\usepackage{subfigure}
\begin{document}

\title[  Detection of transitions in a speech signal
]{%
Detection of  transitions between broad phonetic classes in a speech signal 
}



\preprint{AIP/123-QED}

\author{T V Ananthapadmanabha}

 \email{tva.blr@gmail.com}
 \affiliation{Voice and Speech Systems, Malleswaram, Bangalore}
\author{K V Vijay Girish}
 \email{kv@ee.iisc.ernet.in}

\author{ A G Ramakrishnan}
  \email{ramkiag@ee.iisc.ernet.in}
\affiliation{Indian Institute of Science, Bangalore}%

\date{\today}

\begin{abstract}
Detection of transitions between broad phonetic classes in a speech signal is an important problem which has applications such as landmark detection and segmentation. The proposed hierarchical method detects silence to non-silence transitions, high amplitude (mostly sonorants) to low amplitude (mostly fricatives/affricates/stop bursts) transitions and vice-versa.
 A subset of the extremum (minimum or maximum) samples between every pair of successive zero-crossings is selected above a second pass threshold, from each bandpass filtered speech signal frame. Relative to the mid-point (reference) of a frame, locations of the first and the last extrema lie on either side, if the speech signal belongs to a homogeneous segment; else, both these locations lie on the left or the right side of the reference, indicating a transition frame. When tested on the entire TIMIT database, of the transitions detected, 93.6\% are within a tolerance of 20 ms from the hand labeled boundaries. Sonorant, unvoiced non-sonorant and silence classes and their respective onsets are detected with an accuracy of about 83.5\% for the same tolerance. The results are as good as, and in some respects better than the state-of-the-art methods for similar tasks.
\end{abstract}


\maketitle


\section{Introduction}

\subsection{Segmentation problem}
Speech is perceived as made up of a sequence of discrete sounds. This evokes an expectation that the speech signal can be segmented into a sequence of clearly delineated, non-overlapping intervals corresponding to phones. However, during speech production, the articulators continually move resulting in a speech signal with almost continuous formant tracks. Also, the source process is influenced by the preceding or the succeeding phone as for example the glottal abduction during a vowel-consonant transition, the presence of frication noise following a burst, or the presence of noise components at the onset of a vowel following a strong fricative. Thus, the adjacent phones have a considerable influence on the temporal and spectral properties of a short segment of speech corresponding to the so called current phone\cite{fant}. Hence, speech segmentation is considered a challenging problem. Despite the lack of phone-wise segmentation property in a speech signal, there are clearly marked events or transitions or landmarks arising due to an abrupt change of source process (voiced/unvoiced) and/or an abrupt movement of an articulator (sudden release as in stops, switch over from oral to nasal output as for nasals). Detection of such events serves to guide semi-automatic segmentation or forced alignment or variable frame-rate analysis or analysis around landmarks to extract distinctive features (DFs) or manner classes or phonetic features (PFs).

\subsection{ Literature review}
There are three broad approaches to segmentation: (i) sequential, non-overlapping segmentation based on phones, (ii) parallel, multiple segmentations based on DFs, and (iii) hierarchical segmentation based on PFs. Classification of a speech signal, phone-wise or feature-wise, can also be interpreted as performing segmentation, since it automatically divides the speech signal into distinct segments. 

The first view, viz., the phone based segmentation, is motivated by the perception of speech as discrete units. Accordingly, speech signal is a sequence of non-overlapping intervals, each representing a phone. This approach is widely used in manual labeling of the speech databases. Such a labeling scheme contradicts the acoustic-phonetics knowledge that a given frame of speech signal is strongly influenced by the neighboring phones. However, it is understood that the manual labeling of phones must be considered along with the surrounding context of phones.

In phone level segmentation, the abrupt changes in the short-time spectra are marked as transition events \cite{saijaya}$^-$ 
\cite{ranjani}. Various short-time spectral representations have been used: linear prediction smoothed spectral envelope, ensemble interval histogram, auditory sub-band filter outputs, mel frequency cepstral coefficients (MFCCs), weighted MFCCs, etc. In addition to the standard Euclidean and Mahalanobis distance measures\cite{saijaya}, cross-correlation of short-time spectra\cite{hermert}, model fitting \cite{obrecht}, maximum likelihood estimates and template matching\cite{svendson} have also been used to detect segment boundaries.

Early approach to speech recognition, referred to as template matching approach, attempted segmentation of speech signal and assigned a phone label based on a distance measure between two spectral templates. Such an early approach gives a front-end accuracy of only about 40 to 50\% \cite{cole}. Due to the low front end accuracy, instead of template matching, use of phone dependent cues has been proposed\cite{cole}. However, identification and extraction of such speaker independent cues for all classes of sounds is still an unsolved problem. 

The accuracy of recognition reaches a respectable figure only when alternate pronunciation models are used, with constrained vocabulary and contextual knowledge\cite{harpy}. The need for segmentation is averted using statistical models, such as hidden Markov models, for the classification of phones\cite{jelinek}. It may appear that the statistical approach at once solves both the segmentation and classification problems. The disadvantages with the statistical method are that it requires a huge amount of labeled database for training and it is sensitive to the recording conditions and background noise. Any change in the recording condition may call for re-training. Although hidden Markov model based forced alignment\cite{doroteo} does not require a training database, it assumes that the phone sequence of the utterance to be segmented is known. However, in this paper, we deal with segmentation when the phone sequence is unknown.

The superior speech perception performance of humans in degenerate conditions\cite{lippman} points to the possibility that humans may utilize other sources of knowledge such as distinctive or phonetic features. The importance of integration of phonetic knowledge in speech technology\cite{barry} has been discussed. Several studies have shown that the additional use of DFs or PFs improves the recognition performance of a hidden Markov model classifier\cite{niyogi}$^-$\cite{juneja08}.

The second approach to segmentation is based on distinctive features, a view based on phonology. It is postulated that each speech sound is a bundle of (about 16) binary DFs\cite{jacobson}$^,$\cite{chomsky}. In this model, speech signal consists of a parallel stream of DFs. The presence of each of the DFs extends over different, overlapping intervals with its own boundaries. Acoustic description of DFs given by Jakobson, Halle and Fant\cite{jacobson} remained qualitative in nature as there is no robust automatic method to extract these descriptors. Chomksy and Halle\cite{chomsky} have proposed articulation based DFs. In order to extract these DFs, King and Taylor\cite{king} used a frame-wise analysis with MFCCs and their derivatives (39 parameters) as the acoustic feature vector input to a neural network classifier trained for each DF separately. Though the frame-wise accuracy for the individual DF is high ($>$90\%), the accuracy for the joint or simultaneous occurrence (all correct) of the DFs for a given phone is low (around 50\%), nearly the same as the front-end accuracy of the early template approach. Hidden Markov models\cite{metze} and support vector machine\cite{scharenborg} have also been used for the extraction of DFs. Bromberg et al\cite{bromberg} experimented with a bank of acoustic features along with a bank of classification strategies to extract attributes. No specific feature or classifier uniformly gave a high accuracy. A multilayer perceptron classifier gave an equal error rate (EER) of 10\% for `strident vs silence' but a poor EER of 25\% for the feature `high'. A support vector machine classifier gave an EER of 6\% for `silence' but a poor 42\% for the feature `mid'. 

Some argue that the low scores obtained for extracting articulatory DFs arise since appropriate invariant acoustic features have not been used. According to this view, certain attributes of short-time spectra are invariant for a given place of articulation of stops. That is, the attributes are independent of the influence of coarticulation\cite{blumstein}. However, this view has been questioned\cite{kewly}. A compelling evidence on the existence of invariant attributes is yet to be fully established for all the DFs, though it is an interesting thought.

The third approach, based on the phonetic features, has two models. One of the models is based on the manner and place classification of speech sounds, a view inspired by speech production process. This is similar to the approach of binary DF but with multi-valued DFs and only two parallel streams (manner and place). In their work on DFs, King and Taylor\cite{king} also reported on the identification of manner and place features. The reported frame-wise accuracy is about 90\% for the individual features but the accuracy for all the features being simultaneously correct is only 50\%. A later extension of this study attempted to incorporate mutual dependencies amongst DFs\cite{dbn} and found a marginal improvement in the accuracy. Juneja and Wilson\cite{juneja}$^,$\cite{juneja08} report manner class (silence, vowel, sonorant consonant, fricative and stop) segmentation accuracy of about 79\%\cite{juneja08} on a part of the test set of the TIMIT database, using MFCCs as well as acoustic parameters and support vector machine classifier.

Assuming a bundle of 16 binary DFs, there are $2^{16}=65536$ possible representations and with multi-valued DFs (six manner and eight place), there are $2^{14}=16384$ possible representations, whereas the number of phones is only of the order of 50. This discrepancy arises because a large number of DFs are mutually exclusive. All the DFs may not be relevant for a given phone. For example, the feature `high/low' is relevant only for the manner class `vowels' and irrelevant for consonants and the place feature is irrelevant for vowels. This leads to the second hierarchical classifier model within the approach of phonetic features. Here, a tree structure is used, where each node represents a broad class and is sub-divided into two finer classes\cite{stevens}$^,$\cite{pruthi}. For example, the signal is initially divided into the two broad classes of speech and silence. Then the node `speech' divides into two branches, namely sonorants and non-sonorants. Sonorants are divided into syllabic (vowels) and consonantal (voiced consonants other than stops). The non-sonorants are divided into continuants (fricatives) and interrupted (stops) and so on.

In a hierarchical scheme of PFs, the segmentation problem can be reduced to a set of binary decision making problems. This approach of tackling sub-problems is being pursued by various researchers. We give some select examples, since the number of related publications is very large: (i) onset of a vowel given the segment contains a consonant-vowel transition\cite{prasanna}$^,$\cite{prasanna9} (ii) offset of a vowel given the segment contains a vowel-consonant transition\cite{sreenivasa} (iii) semi-vowels\cite{epsysemivowel}, laterals\cite{epsylateral}, nasals\cite{pruthi} (iv) fricatives\cite{alifricative} (v) stop bursts\cite{bitar}$^,$\cite{apr} (vi) manner classes\cite{hopkins} (vii) trills\cite{bhaskar}. Alternately, if the phone corresponding to the final terminal is determined, then all the higher level nodes leading to that branch are also determined. For example, if a stop is detected from a segment of a continuous speech signal, then that segment is automatically assigned all the higher level nodes, viz., interrupted, non-sonorant, speech. Hence, research at determining an individual phone, irrespective of context, is also being pursued. 

An issue related to the extraction of phonetic features is the landmark detection, landmark being an important transition dividing a speech signal into certain broad segments\cite{stevens}$^,$\cite{saloman}$^,$\cite{liu}. Conventionally, speech analysis for the extraction of acoustic features is carried out frame-wise. But in this alternative approach, speech signal around the landmarks is analyzed to extract the acoustic features, which are subsequently given as an input to a classifier to determine either the phones or phonetic features. Liu\cite{liu} has used the change of energy, over six sub-band signals, between two frames spaced 50 ms apart, for detecting four broadly defined landmarks. Salomon et al\cite{saloman} have used a set of twelve temporal parameters to detect three landmarks as well as for manner classification.

\subsection{ About this work}

Reddy\cite{reddy} proposed the use of intensity differences (peaks and valleys) to detect certain broad classes of sounds for a limited vocabulary, speaker dependent task. Stevens\cite{stevens} has observed that certain landmarks may be located based only on abrupt amplitude changes in a speech signal. This paper proposes four different measures for detecting transitions between broad phonetic classes in a speech signal. 

A measure is defined on the quantized speech signal to detect transitions between very low amplitude or silence (S) and non-silence (N) segments. These very low amplitude regions could be stop closures, pauses or silence regions at the beginning and/or ending of an utterance. 

We propose two other measures to detect the transitions between relatively high (H) and low (L) amplitude segments and vice-versa. We make use of the fact that most sonorants have higher energy in the low frequencies, than other phone classes such as unvoiced fricatives, affricates and unvoiced stops. For this reason, we consider a bandpass speech signal below 500 Hz for processing. For a transition within a sonorant (vowel to voiced consonant or vice-versa), the amplitude of a bandpass filtered speech signal will nearly be the same. However, for a transition from a sonorant to any of the unvoiced consonants, the amplitude changes suddenly from a high to a low level across the transition. On the other hand, when there is a transition from a stop consonant (or fricative) to a vowel, the amplitude suddenly increases from a low to high level at the vowel onset. Thus, by tracking the locations of relative peak (valley) amplitudes in successive closely spaced analysis frames, we can detect the transitions between the broad phonetic classes in a speech signal. 

When the amplitude of the bandpass filtered signal in a frame is very low, then a sudden change in the level cannot be assumed to represent a transition between phonetic classes, with any kind of certainty. To avoid such spurious insertions, we introduce the fourth measure, viz., the mean peak to valley amplitudes for a frame. When this measure is very low, any transition within such a frame is ignored.

The above rationale for the selection of features and the proposed algorithm for detecting transitions is based on an expectation born of the knowledge of the different phonetic classes of speech.

Combining the above types of transitions, the speech signal is divided into the five broad homogeneous classes: silence (S), high (H), low (L), high-low (HL) and low-high (LH). Based on homogeneous classes, the speech signal is classified into the broad phonetic classes of sonorants, unvoiced sonorants, bursts and silence. The proposed method is validated using the entire TIMIT database. The accuracy of detection and the temporal accuracy of the onset of these classes are computed. The results are noted to be comparable to those of state-of-the-art methods.

\section{PROPOSED TEMPORAL MEASURES}

\subsection{Silence Index}

For detecting the transitions between silence and non-silence classes, the speech utterance $s[n]$ is demeaned and then normalized as:
\begin{equation}
s_z[n]=s[n]-\dfrac{\sum_{1}^{M}s[n]}{M} \;; s_{N}[n]=\dfrac{s_z[n]}{\max |s_z[n]|}
\label{normsp}
\end{equation} 
where $M$ is total number of samples in the utterance. 

Conventionally, if the energy of a frame is very low compared to the maximum normalized speech signal energy, the frame is considered to be silence. Also, the peak amplitude within a frame relative to the global peak amplitude is sometimes used to identify silence frames. We define a new measure, silence index, for silence detection. The normalized speech signal of 16 bits resolution is quantized to 9 bits (the least significant 7 bits are zeroed) and a staircase signal is obtained. The size of the analysis frame is 10 ms. The number of successive samples having the absolute value within  one least significant bit is counted. During every zero crossing, it is likely that there are two samples close to zero and within the least significant bit. To avoid these samples being counted as silence samples, only when there are a minimum of three successive samples below threshold, they are counted. 

Silence Index (SI) is a ratio, defined as

\begin{equation}
    SI=\dfrac{count\; of\; samples\; below\; threshold}{number\; of\; samples\: in\: the\: frame }
    \end{equation}
The value of SI is assigned as a feature of the mid 5 ms segment of the frame. Since a frame shift of 5 ms is used, there is a new value of SI for every 5 ms segment.

\begin{figure*}
\mbox{
\includegraphics[width=.31\textwidth,height=.22\textheight]{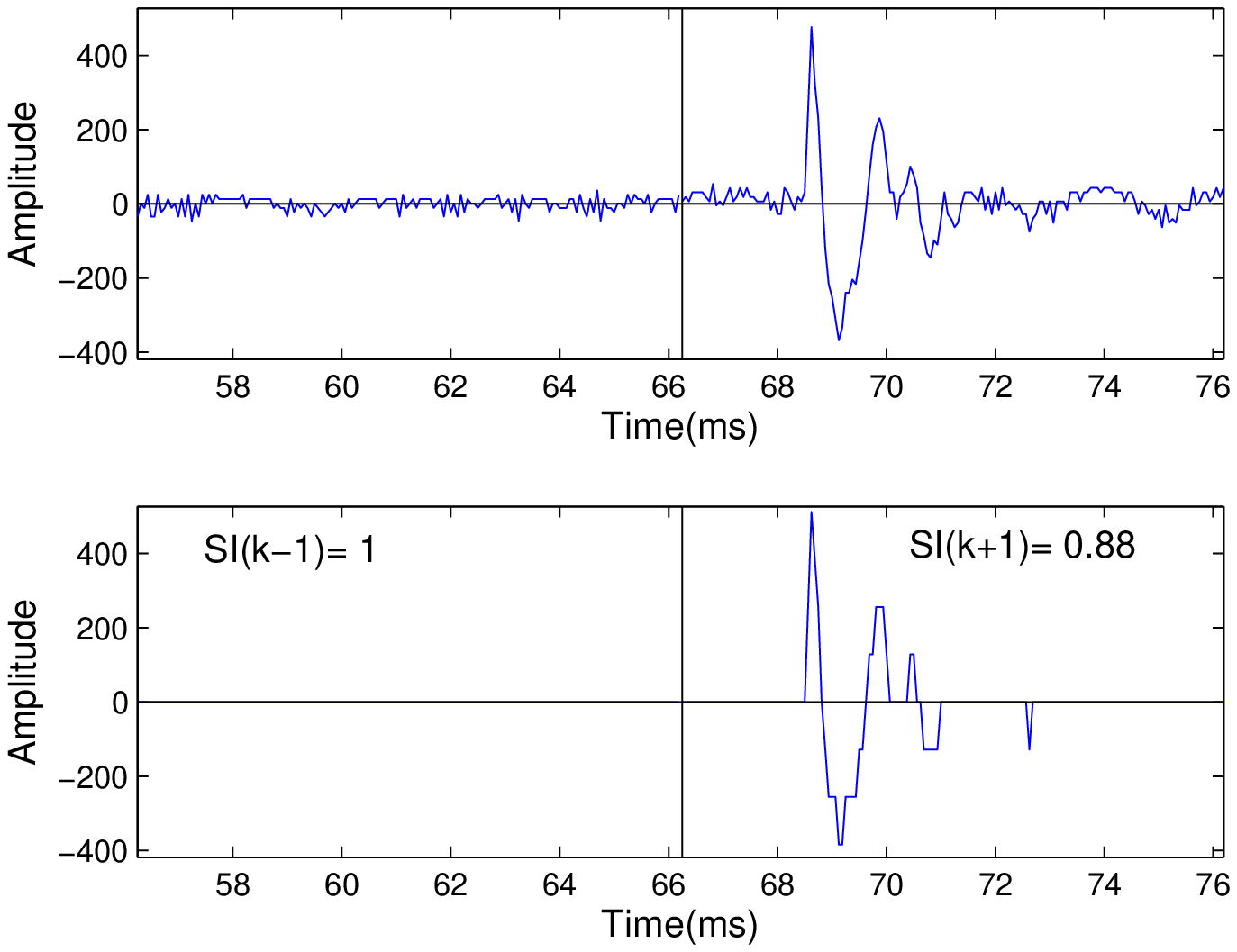}

\includegraphics[width=.31\textwidth,height=.22\textheight]{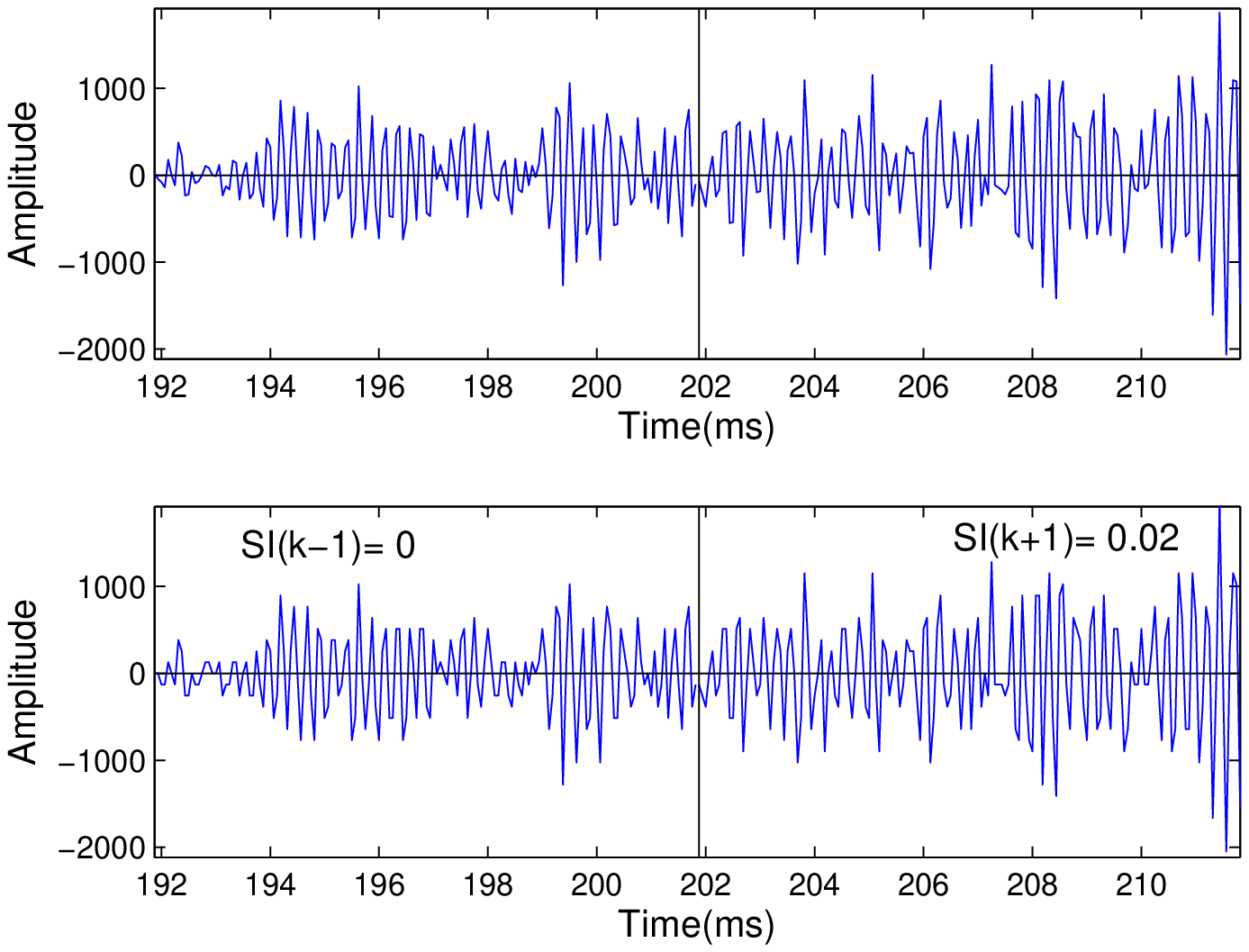}

\includegraphics[width=.31\textwidth,height=.22\textheight]{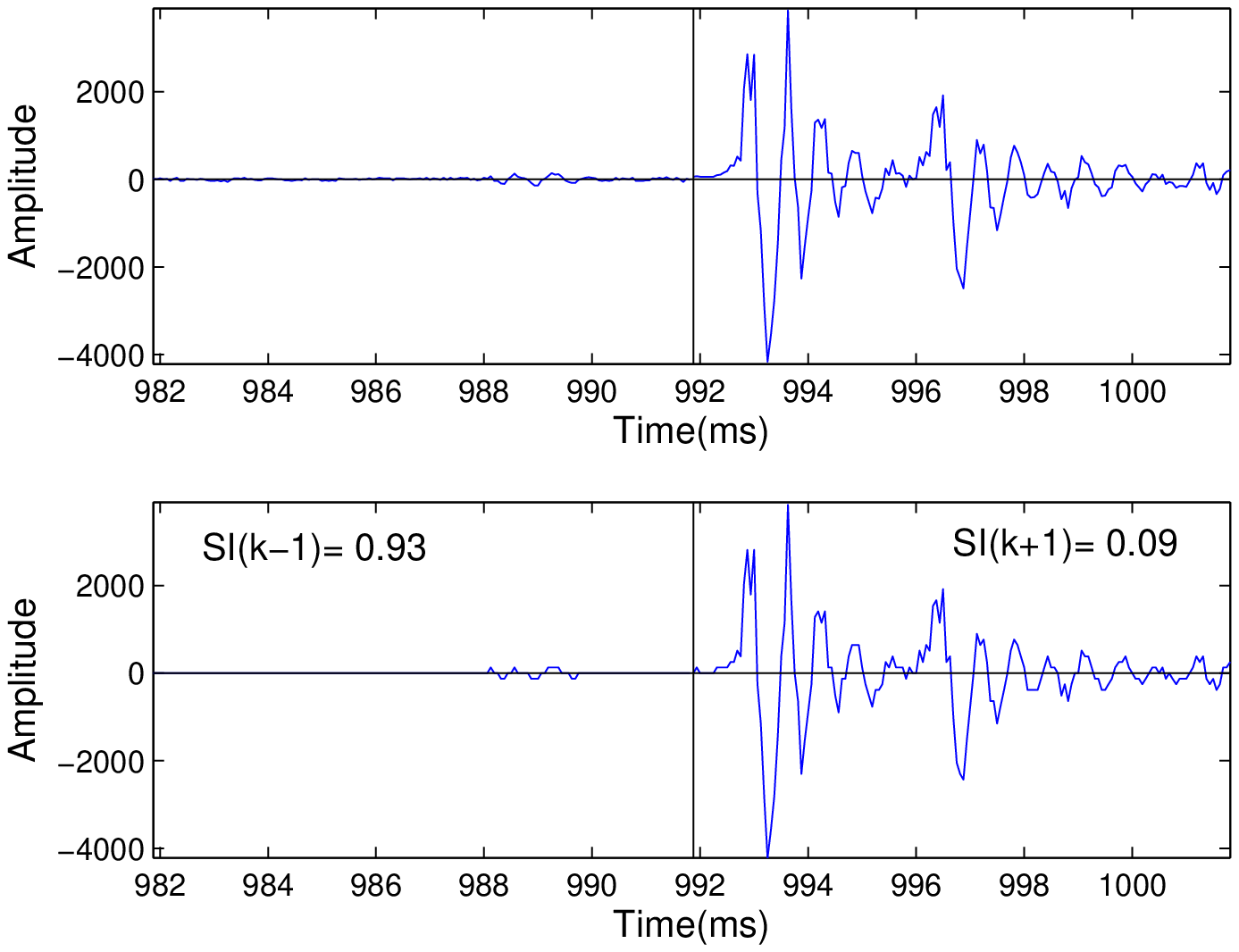}
}
\caption{Variation of silence index (SI) values with the variation in the nature of the signal across consecutive frames and the corresponding quantized signals. (a) Presence of a high amplitude pulse in a silence segment. (b) An unvoiced segment. (c) A stop closure to burst transition.}
\label{sivus}
\end{figure*}

Figure \ref{sivus} shows the signal and its corresponding quantized counterpart, together with the SI values for three types of speech segments, each containing three overlapping frames: (a) silence followed by an impulse, (b) unvoiced and (c) a closure to burst transition segment. It may be noted that SI has a very high value for the silence segment even in the presence of a large amplitude impulse. The SI is low for the unvoiced segment. During a closure-burst transition, there is a sharp decrease in the value of SI for two successive frames. We make use of such abrupt changes in the value of SI for detecting the transitions from/to silence segments.

\subsection{Features based on the extrema in a frame}



For detecting the transitions between low and high amplitude classes, the normalized speech signal is bandpass filtered (BPF) using the bell cosine shaped filter in the frequency domain:

\[ \hspace{-.0cm}
h_p[f]=\left\{\begin{array}{l }
0.5-0.5\cos\left(\pi\left(\frac{f-f_1/2}{f_1/2}\right)\right),\; f_1/2\leq f< f_1\\
1\hspace{3.5cm} ,\;f_1\leq f \leq f_2/2\\
0.5+0.5\cos\left(\pi\left(\frac{f-f_2/2}{f_2/2}\right)\right),\; f_2/2< f\leq f_2\\
0,\;\; elsewhere\\
\end{array}\right\}
\]
$f_1=70 \;$Hz and $f_2=500\;$ Hz. This filter has a cosine rising function from 35 to 70 Hz, unit gain from 70 to 250 Hz and cosine falling function from 250 to 500 Hz. The 3 dB frequencies of the bandpass filter are 60 and 340 Hz. This is close to `Band 1' used by Liu\cite{liu} for landmark detection. Instead of computing energy difference across 50 ms\cite{liu}, we have defined new temporal measures. The BPF signal $s_{B}$ is analyzed with a longer frame size of 40 ms, twice the duration of the worst case maximum pitch period of 20 ms, corresponding to a pitch frequency of 50 Hz.

\subsubsection{Selection of extrema based on a dynamic two pass threshold}
Speech signal is non-stationary and has a rapid variation in the time domain. So, the extraction of the features from $s_{B}[n]$ needs to adapt to the variation in the signal from frame to frame. Let $s^j_{B}$ denote the bandpass filtered signal between the first and the last zero crossings in the $j^{th}$ frame. We define features based only on those extrema in $s^j_{B}$ obtained with a 2-pass, frame adaptive threshold. 

The first pass positive threshold $T^j_{P1}$ is defined as
 \begin{equation}
	T^j_{P1}=mean(\{ s^j_{B}[n] \})\;\forall  s^j_{B}[n]>0,
\end{equation}

From $s^j_{B}$, all the positive maxima $p^j_{B}$ between successive zero crossings are obtained. A subset of these maxima is selected as: 
\begin{equation}
 p^j_{B1}=\{p^j_{B},\; s.t.\;  p^j_{B}>0.5 T^j_{P1} \}
 \end{equation}  
 
The second pass positive threshold, $T^j_{P2}$ is defined as :
\begin{equation}
T^j_{P2}=mean\{p^j_{B1}\}
\end{equation}
The set of maxima after the second pass is obtained as
\begin{equation}
p^j_{B2}=\{p^j_{B1},\: s.t.\; p^j_{B1}\geq 0.5T^j_{B2}\}
\end{equation}
Note that a factor of 0.5 is applied on the threshold in order to capture the maxima in the rapidly changing voiced segments. 
Similarly, from the valleys (negative peaks) between successive zero crossings, the set of minima $v^j_{B2}$ is obtained. 

\subsubsection{Relative positions of first and last extrema in a frame}
The mid-sample of each frame is used as the relative reference and the positions of the first extremum (PFE) and the last extremum (PLE) in $p^j_{B2}$ or $v^j_{B2}$ are measured with respect to this reference sample. Thus, a location to the left of the reference is considered negative. The values of PFE and PLE are treated as the features of the mid 5 ms segment of the frame.

Example: Figure \ref{voicunv}(a) shows a segment of  voiced speech signal and the corresponding bandpass filtered signal. Though there is a change in the signal structure, the entire signal within the frame belongs to a homogeneous class. The positive and negative second pass thresholds are shown by the horizontal lines above and below zero, respectively. The extrema after second level thresholding are also shown, along with the PFE and PLE, which lie on either side, far away from the reference instant i.e. $PFE\ll 0$ and $PLE\gg0$. Since only the segment of signal within the frame between the first and last zero crossings is considered for analysis, the first maximum around 0.5 ms, which lies to the left of the first zero-crossing, is not considered. In this example, the extremum for PFE is a negative peak whereas that for PLE is a positive peak. Thus, the measurement of PFE and PLE does not take into account the polarity of the extremum.

Figure \ref{voicunv}(b) is similar to \ref{voicunv}(a) but for an unvoiced speech segment. Once again, we notice that the extrema corresponding to PFE and PLE lie on either side, far away from the reference instant of the frame i.e. $PFE\ll 0$ and $PLE\gg0$. Thus, this property of $PFE\ll 0$ and $PLE\gg0$ is satisfied whenever the speech signal within a frame corresponds to a homogeneous class.

Figure \ref{voicunv}(c) is similar to \ref{voicunv}(a) but for a transition from a voiced to an unvoiced segment. Note that both PFE and PLE are negative. The converse property of both PFE and PLE being greater than zero is satisfied for an unvoiced to voiced transition shown in Fig. \ref{voicunv}(d). 

From the above illustrations, we can deduce the following:
(a) When PFE and PLE have opposite signs, the frame corresponds to a homogeneous class. (b) $PFE \ll 0$ and $PLE \le 0$ for a typical case of a transition from a high to a low amplitude (H-L) segment. (c ) $PFE \ge 0$ and $PLE \gg 0$ for a typical case of a transition from a low to high amplitude (L-H) segment.

Thus, we can divide the speech signal into homogeneous H-class or L-class (PFE and PLE having opposite signs) and the two types of transitions H-L and L-H (PFE and PLE having the same signs).  

\subsubsection{Mean absolute difference of extrema in a frame}

\begin{figure*}

\centering
\mbox{\includegraphics[width=.5\textwidth,height=.22\textheight]{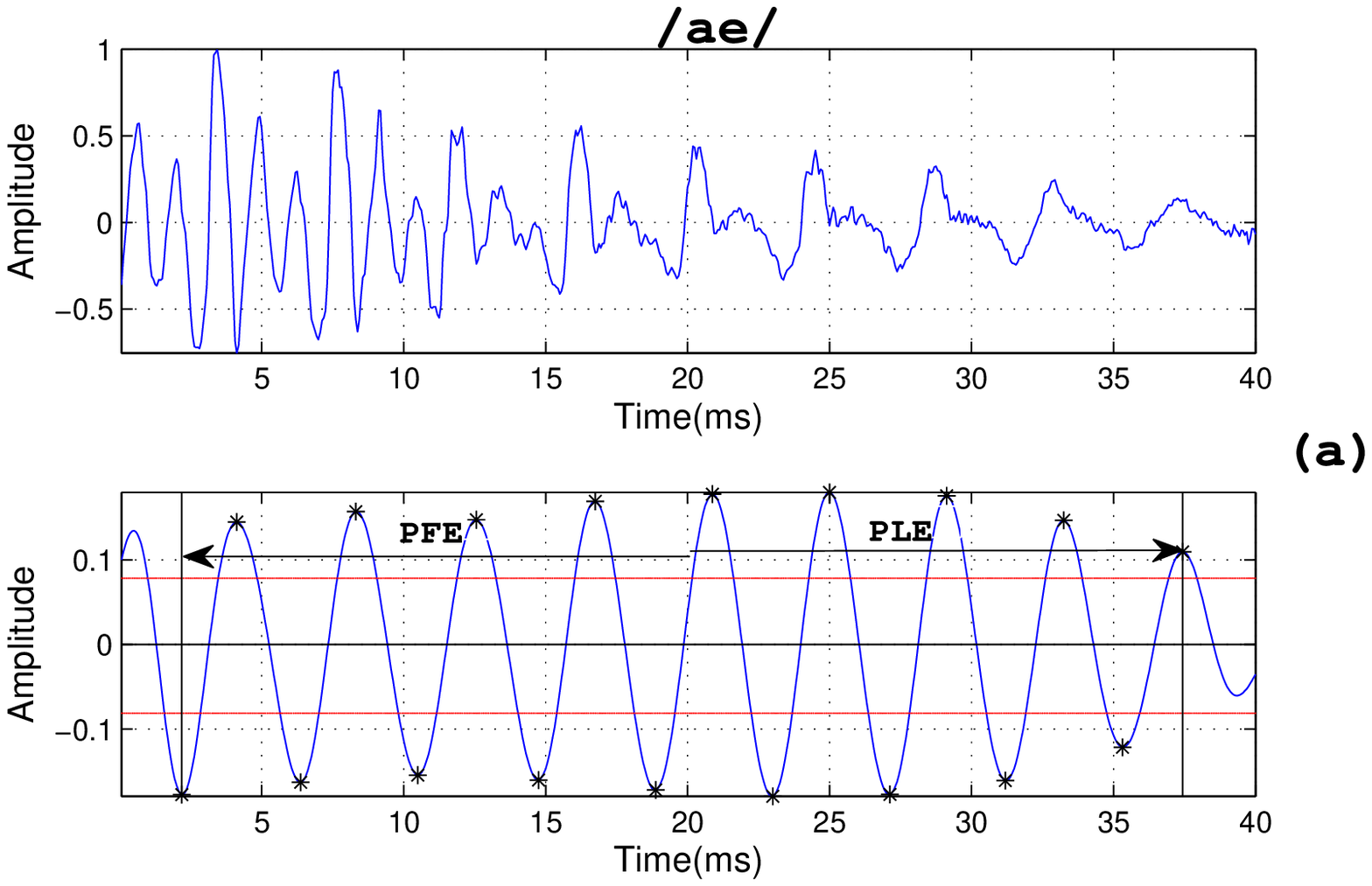}

\includegraphics[width=.5\textwidth,height=.22\textheight]{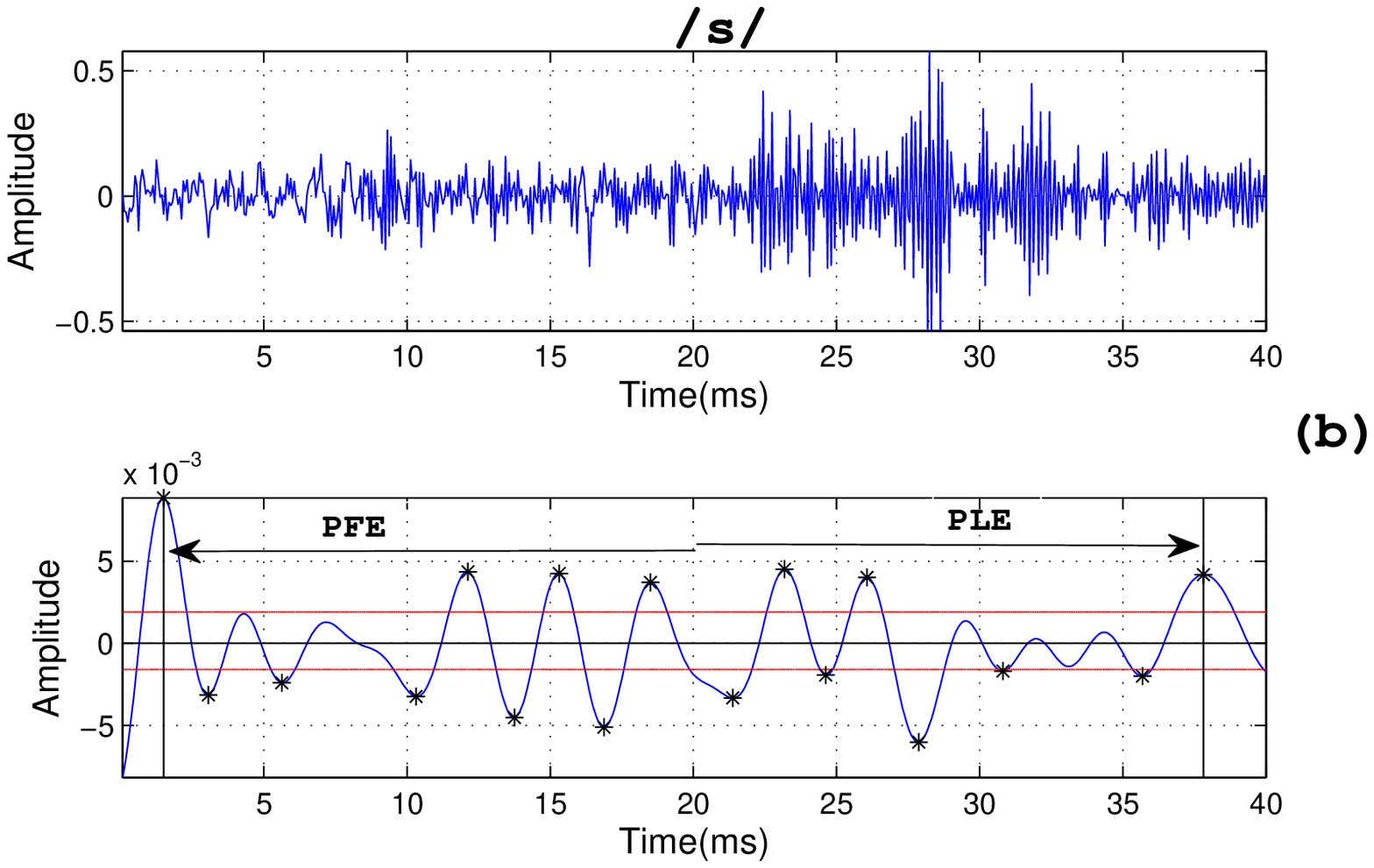}}
\mbox{

\includegraphics[width=.5\textwidth,height=.22\textheight]{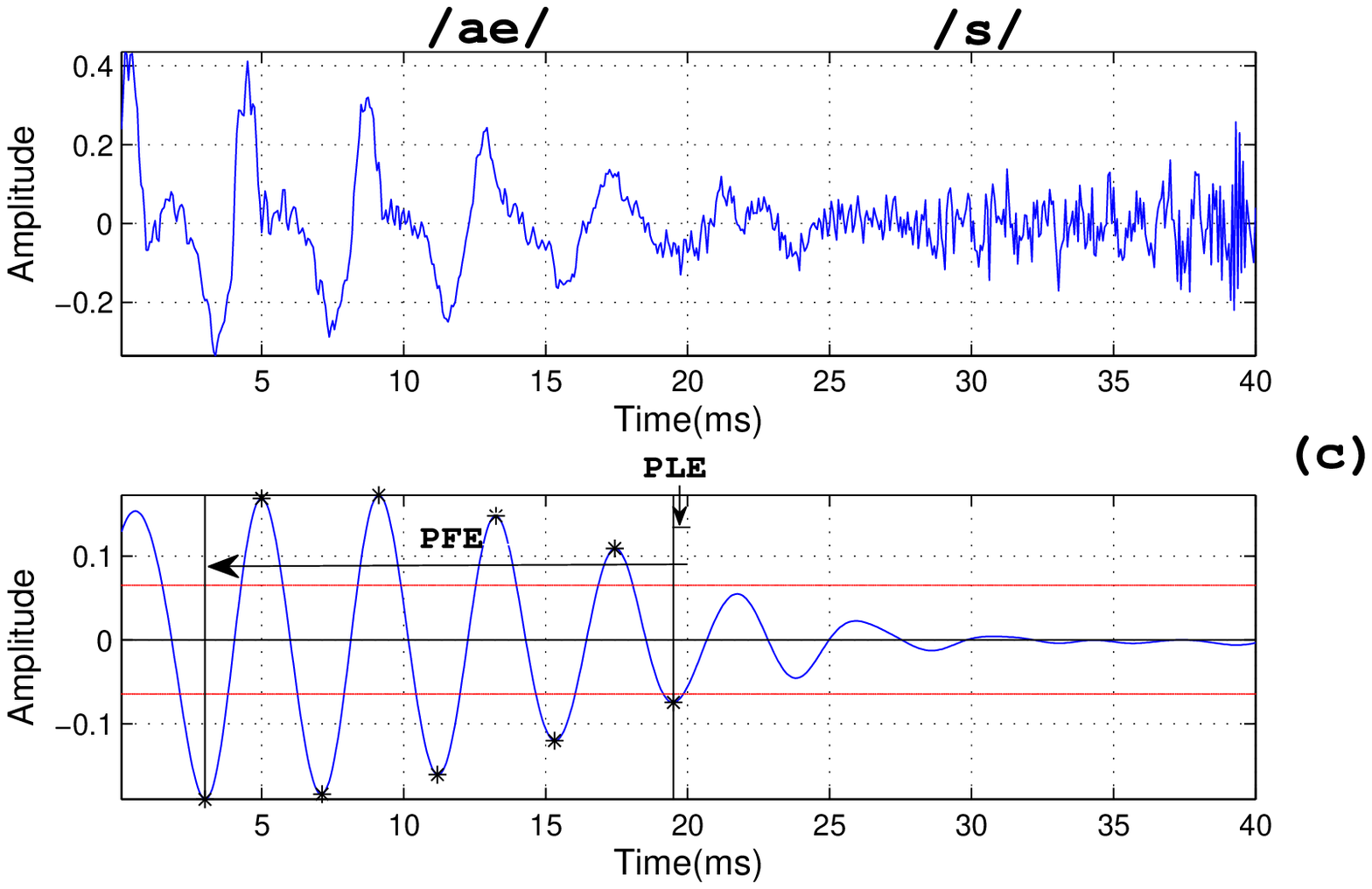}

\includegraphics[width=.5\textwidth,height=.22\textheight]{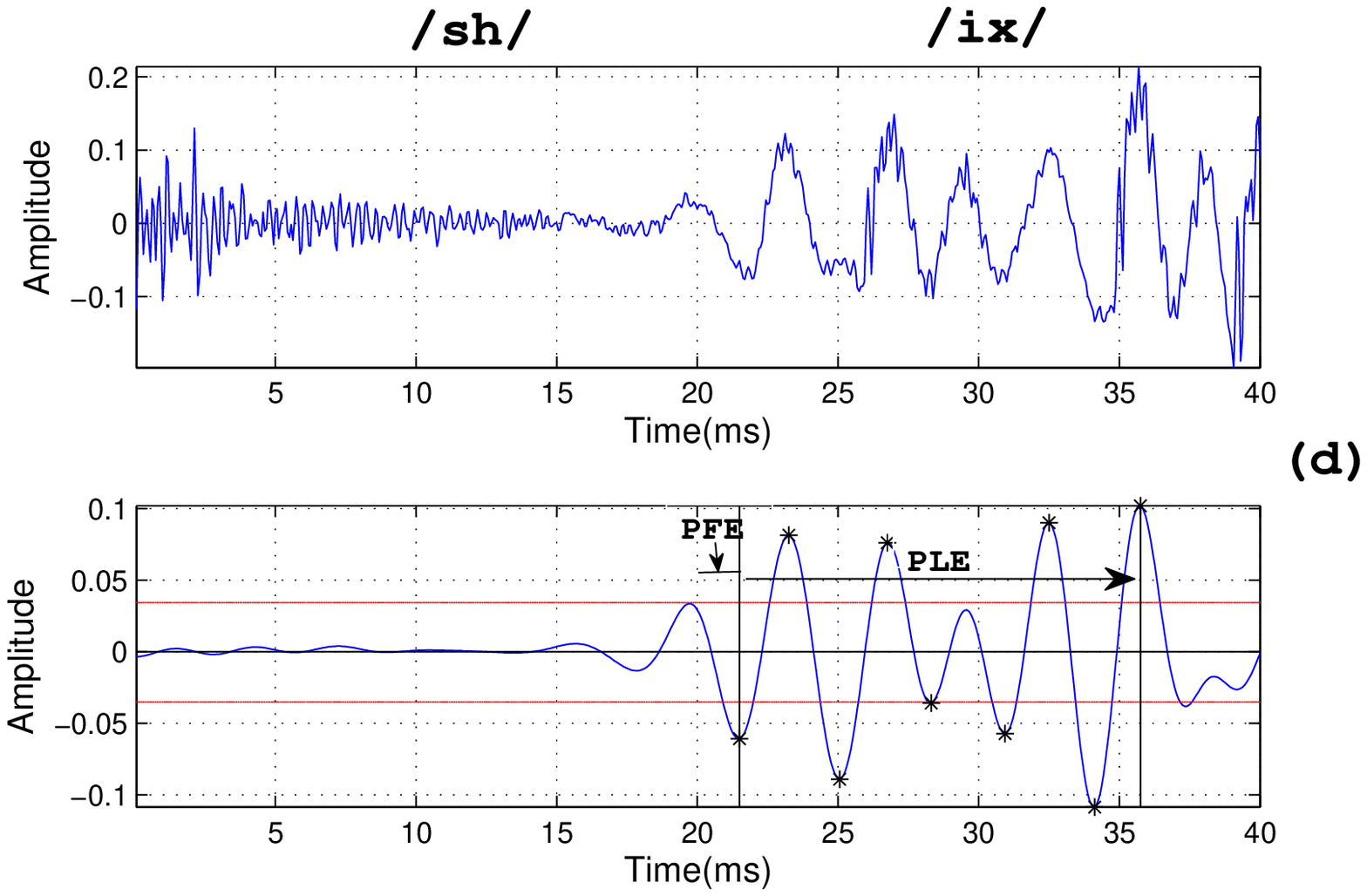}
}
\caption{Sample frames of speech signal and their corresponding bandpass filtered versions. The extrema above the second pass thresholds (horizontal lines above and below zero) as well as the positions of the first (PFE) and last extrema (PLE) are shown. (a) A voiced segment (ADE= 0.32). (b) An unvoiced segment  (ADE= 0.01). (c) A voiced-unvoiced transition (ADE= 0.31). (d) An unvoiced-voiced transition (ADE= 0.27). }
\label{voicunv}
\end{figure*}

Though the peak value of BPF signal in Fig.\ref{voicunv}(b) is very low, of the order of 0.005, it is still shown as an example of a homogeneous unvoiced segment. Any transition based on PFE or PLE is considered as unreliable when the level of BPF signal is very low. For this purpose, another measure named as absolute difference of extrema (ADE) is introduced, which is the mean value of the absolute difference between positive and negative extrema. The caption for  Fig.\ref{voicunv} also gives the values of ADE for each of the sample signals shown. When ADE is below a threshold (0.02), any transition due to PFE or PLE is ignored as unreliable. Thus, spurious insertions of transition in unvoiced segments and frication following bursts having sporadic spike like signal components are avoided by this rule. A transition during a silence to unvoiced phone or burst would also be missed as the entire segment will have ADE less than 0.02. However, an abrupt change in silence index captures these silence to non-silence transitions such as closure to burst and pause to fricative.

\section{Algorithm for detection of transitions}
\begin{figure}
 
\centering
 \includegraphics[width=.45\textwidth,height=.23\textheight]{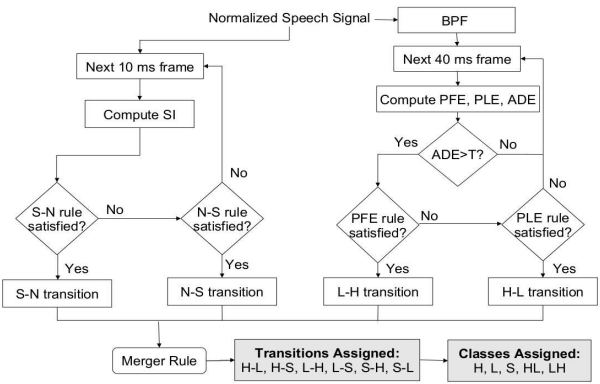}

\caption{Flowchart showing the detection of transitions and the assigned classes. (T is the threshold for ADE).}
 \label{flwchrt}
 \end{figure}

We refer to the proposed algorithm shown in Fig.\ref{flwchrt} as AGR algorithm. The rules cited in the flowchart are presented in Tables \ref{sirule}-\ref{clasmerger}. We discuss the strategy used in the algorithm with an example. In a hierarchical classifier, the first step is to divide the speech signal into silence and speech segments. Hence, we first present the detection of transitions from/to silence.

\subsection{\label{sn}Detection of transitions between silence and non-silence classes}

The characteristic changes in the value of SI for three consecutive frames are used to detect the transition from silence (S) to non-silence (N) classes or vice-versa from the speech signal. These rules are listed in Table \ref{sirule}. The actual instant of transition is determined by recomputing SI for non-overlapping sub-segments of 1 ms duration. The instant of transition is defined as the cross-over point when SI crosses the value of 0.5 in any direction. The samples between an N-S and a following S-N transitions are labeled as S-class and vice versa, as shown in Table \ref{succtrans}.

SI for a speech segment extending over several phones and the S-N and N-S transitions detected using SI are shown in Fig.\ref{transs}(a). The silence to non-silence transitions around 200 ms corresponding to the boundary between `h\#' and /sh/ and around 770 ms corresponding to the boundary between /dcl/ and burst /d/ are detected successfully. However, the /dcl/ segment (550-580 ms) between /eh/ and /jh/ is missed as considerable energy is present in the corresponding segment. The phone /jh/, a voiced affricate, is realized as unvoiced in this utterance. It is not clear if a closure needs to necessarily be marked for an affricate realized as a fricative.  
 
   \begin{table}
 
       \caption{Rules for detecting transitions between silence (S) and non-silence (N) classes. }
             \label{sirule}
       \begin{ruledtabular}
       \begin{tabular}{lc}
      
              SI values and the past context & Type \\\hline
                
       $SI(k-1)\geq0.6 $ and $SI(k+1)\leq0.4 $ & S-N\\
       $SI(k)\leq0.35$ and $Previous\; transition=$N-S &S-N\\
    $SI(k-1)\leq0.4 $ and $SI(k+1)\geq0.6 $  & N-S\\
     $SI(k)\geq 0.7$ and $Previous \; transition=$S-N & N-S\\
      
       \end{tabular}
       \end{ruledtabular}
       \end{table}
    
\subsection{Detection of H-L and L-H  transitions }
Table \ref{PFErule} shows the rules for detecting L-H and H-L transitions. Appropriate class labels need to be assigned to the segments between successive transitions. For example, in the simplest case, a segment which lies between H-L and L-H transitions would naturally be labeled as L-class. The labelings used are shown in Table \ref{succtrans}.
  
A segment of speech signal and its bandpass version are shown in Figs.\ref{transs}(a) and \ref{transs}(b), rspectively. The values of PFE and PLE obtained for successive frames are scaled and plotted as a function of time in Fig. \ref{transs}(b). Towards the end of /sh/, just before 300 ms, PFE rapidly increases from a negative to a positive value and returns to a negative value for the next phone. The positive to negative zero crossing (NZC) in PFE marks a strong L-H transition (first row of Table \ref{PFErule}). We choose the zero-crossing in the BPF signal closest to this NZC as the transition instant.
 
The homogeneous segment corresponding to /ix hv eh dcl/ all belong to H-class with PFE and PLE having opposite signs. In this example, the segment labeled as /dcl/, if valid, is  expected to be in S class but happens to be in H-class since it has a relatively large amplitude. Towards the end of /eh dcl/ and approaching /jh/, since there is a relatively large change from high amplitude to low amplitude, PLE rapidly falls and changes from a positive to a negative (NZC) and back to a positive value. This change in PLE marks a strong H-L transition (third row of Table \ref{PFErule}). We choose the zero-crossing in the BPF signal closest to NZC as the transition instant.

During the H-L transition from /ih/ to /dcl/ around 720 ms, there is an abrupt decrease in amplitude (unlike /eh/ to /dcl/). PLE decreases rapidly to a minimum, close to the base line, without a sign change. If this minimum value of PLE is within 5 ms, then it is considered a weak H-L transition (fourth row of Table \ref{PFErule}). A similar weak L-H transition (second row of Table \ref{PFErule}) due to PFE is seen between /dcl d/ and /ah/ around 800 ms. We consider a weak transition also as a genuine transition. The distinction between a strong and weak transition is noted only for the sake of further analysis, if required.

It is not necessary that L and H classes always alternate. It may be noted that across /kcl-k/ and /s-ux/, there are two consecutive L-H transitions due to PFE. In order to distinguish such transitions, the segment between two consecutive L-H transitions is denoted as a HL class. Similarly the signal between two consecutive H-L transitions is labeled as a LH class. Such occurrences of LH and HL classes are rare. However, this specific example of HL class is an exception. In this example, though the segment /k s/ should have been labeled HL class, the first transition across /kcl-k/ due to PFE is ignored since ADE is below threshold and hence the label happens to be L. The transition /kcl k/ is still captured as S-N transition based on SI as shown in Fig.\ref{transs}(a). The class label of a speech segment may change when the characteristics of SI and PFE/PLE are combined by the merger rule described below.

         \begin{table}
       \caption{Rules for detecting transitions based on contours of PFE and PLE. NZC, MIN and MAX denote a positive to negative zero crossing, a local minimum and maximum, respectively of PFE or PLE contour.}
       \label{PFErule}
       \begin{ruledtabular}
       \begin{tabular}{lc}
      
Nature of PFE and PLE in the frame & Type \\\hline
                
      NZC in $PFE$, $PLE\gg0 $ & Strong L-H\\
    PFE is a MAX, $ |PFE| \leq 5\:ms$, $PLE\gg0 $   & Weak L-H\\
      $PFE\ll0 $, NZC in $PLE$ & Strong H-L\\
      $PFE\ll0 $, PLE is a MIN, $ |PLE| \leq 5\:ms$, & Weak H-L\\

       \end{tabular}
       \end{ruledtabular}
       \end{table}
       
   \begin{table}

   \caption{Class labeling of a segment between successive  transitions.}
      \label{succtrans}

   \begin{ruledtabular}
   \begin{tabular}{ccc}
  
          \multicolumn{2}{c}{Types of successive transitions} & Class label assigned\\\hline
             $k^{th}$& $(k+1)^{th}$&\\ \hline  

   N-S & S-N& S\\
    S-N & N-S& N\\ \hline
   L-H & H-L& H\\
   H-L& L-H & L\\
   H-L & H-L& LH\\
   L-H & L-H & HL\\
   
   \end{tabular}
   \end{ruledtabular}
   \end{table}
   
\subsection{Merger rule and class assignment}
SI is computed with a frame duration of 10 ms
on the speech signal whereas the frame duration used for the measurement of PFE
and PLE is 40 ms and it is obtained from the BPF signal.
 Thus, PFE and PLE are computed independently of SI. The transitions are also marked independently using these measures. However, it is ensured that the decisions assigned by the two processes correspond to the same 5 ms inter-frame segment. The very first frame of any utterance is assumed to be a silence frame. After detecting the transitions, their locations are arranged in an ascending order of sample index (time). It is highly likely that a detected transition due to PFE or PLE may lie close to a transition detected using SI. For example, a vowel offset followed by a stop closure would be detected as a H-L and also as a N-S transition; a silence followed by a vowel onset would be detected as a L-H as well as a S-N transition. Such cases of redundant (duplicate) transitions have to be merged into a single transition. Hence, decisions have to be made on the temporal spacing allowed between the two types of transitions to merge them into one and the same transition and on the location of the new, merged transition. These decisions are called the merger rules.

The rules for merging the two types of transitions are listed in Table \ref{merger}. 

(a) Any H-L or L-H transition detected within a silence class is marked for removal and the type of transition is noted for N-class assignment. This obvious condition is not shown in the Table.

(b) When a S-N transition is followed by a L-H transition within 10 ms or by a H-L transition within 20 ms, then the latter transition $(k+1)^{th}$ is marked for removal. As a consequence, the $(k+2)^{th}$ transition will now become $(k+1)^{th}$ transition. 

(c) When a H-L or L-H transition is followed by a N-S transition within 20 ms, then the former transition is marked for removal. 

Table \ref{clasmerger} lists the rules for assigning class labels to the segments lying between successive transitions.

    \begin{table}
    \caption{Rules for merging redundant transition labels.}
    \label{merger}
    \begin{ruledtabular}
    \begin{tabular}{cccc}
   
    \multicolumn{2}{c}{Successive Transitions}& Separation (ms) & Outcome\\\hline
    $k^{th}$& $(k+1)^{th}$& &\\ \hline
    S-N& L-H & $\leq 10$ ms & Remove $(k+1)^{th}$\\
    S-N& H-L & $\leq 20$ ms & Remove $(k+1)^{th}$\\
       L-H & N-S & $\leq 20$ ms & Remove $k^{th}$\\
           H-L & N-S & $\leq 20$ ms & Remove $k^{th}$\\
    \end{tabular}
    \end{ruledtabular}
    \end{table}
    
        \begin{table}
        \caption{Class assignment after the application of the merger rules. k and k+1 denote the revised frame indices.}
        \label{clasmerger}
        \begin{ruledtabular}
        \begin{tabular}{ccc}
       
        \multicolumn{2}{c}{Types of successive Transitions} & Class assigned\\\hline
           $k^{th}$& $(k+1)^{th}$&\\ \hline
        S-N& L-H& L\\
        S-N& H-L& H\\
        L-H& N-S & H\\
        H-L& N-S& L\\
           
        \end{tabular}
        \end{ruledtabular}
        \end{table}

The segment between a S-N and the next transition (after the above merger step) is labeled as L or H-class, respectively, depending upon whether the next transition is of type L-H or H-L. If the next transition is of type N-S (because of a H-L or L-H transition that was removed), then either of the class labels H or L which occupied most of the duration between S-N and N-S is assigned to the segment (not shown in Table \ref{clasmerger}).
The segment before a N-S transition is labeled as H- or L-class if the preceding transition is a L-H or H-L transition, respectively. If the preceding transition is of type S-N (because of a removal), it is handled similar to the situation discussed above.

Thus, the following five classes result after both the sets of rules are applied on the transitions: (a) H (b) L (c) S (d) HL (f) LH.
 
\begin{figure*}

\includegraphics[width=.75\textwidth,height=.22\textheight]{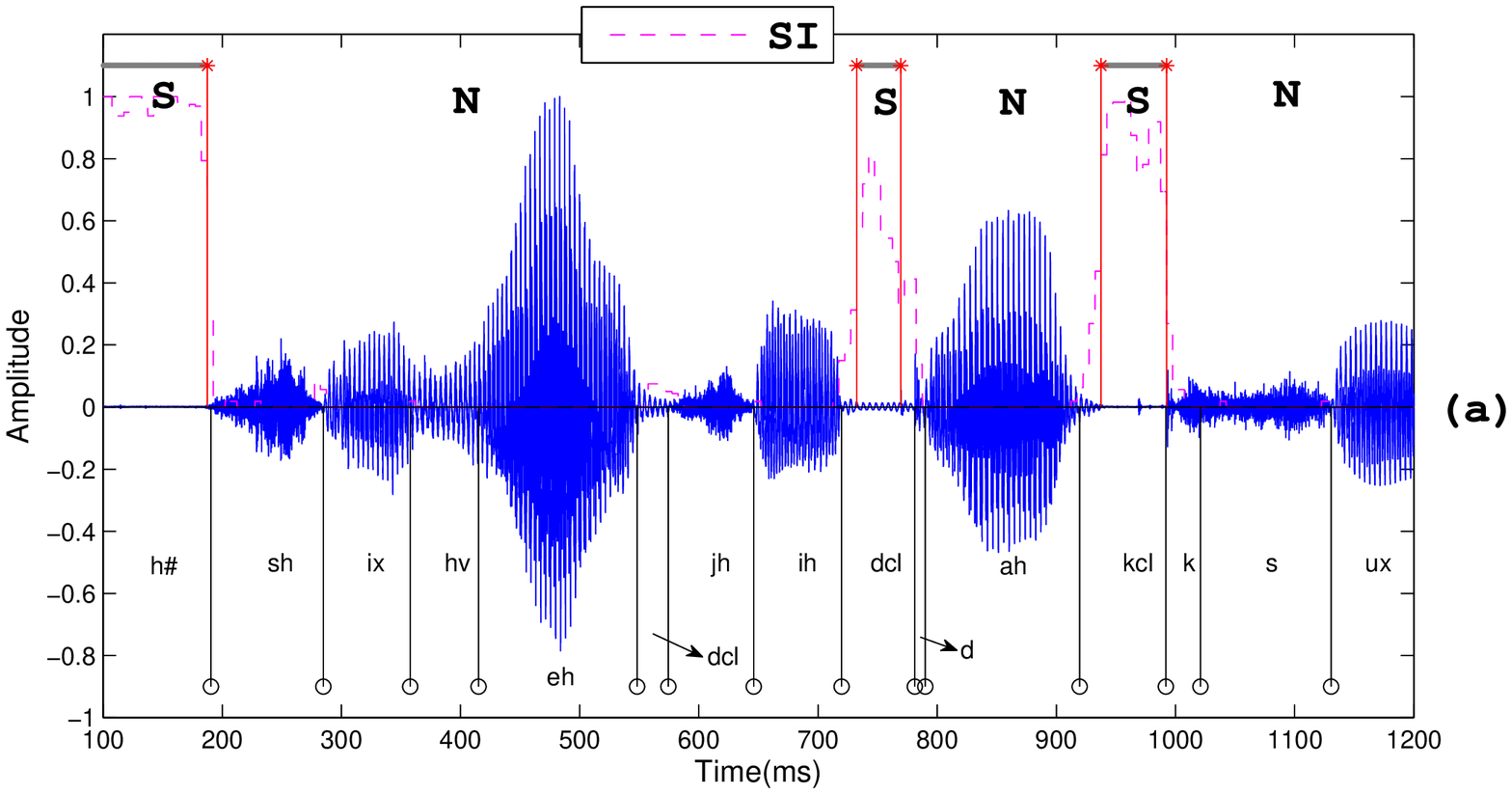}

\includegraphics[width=.75\textwidth,height=.22\textheight]{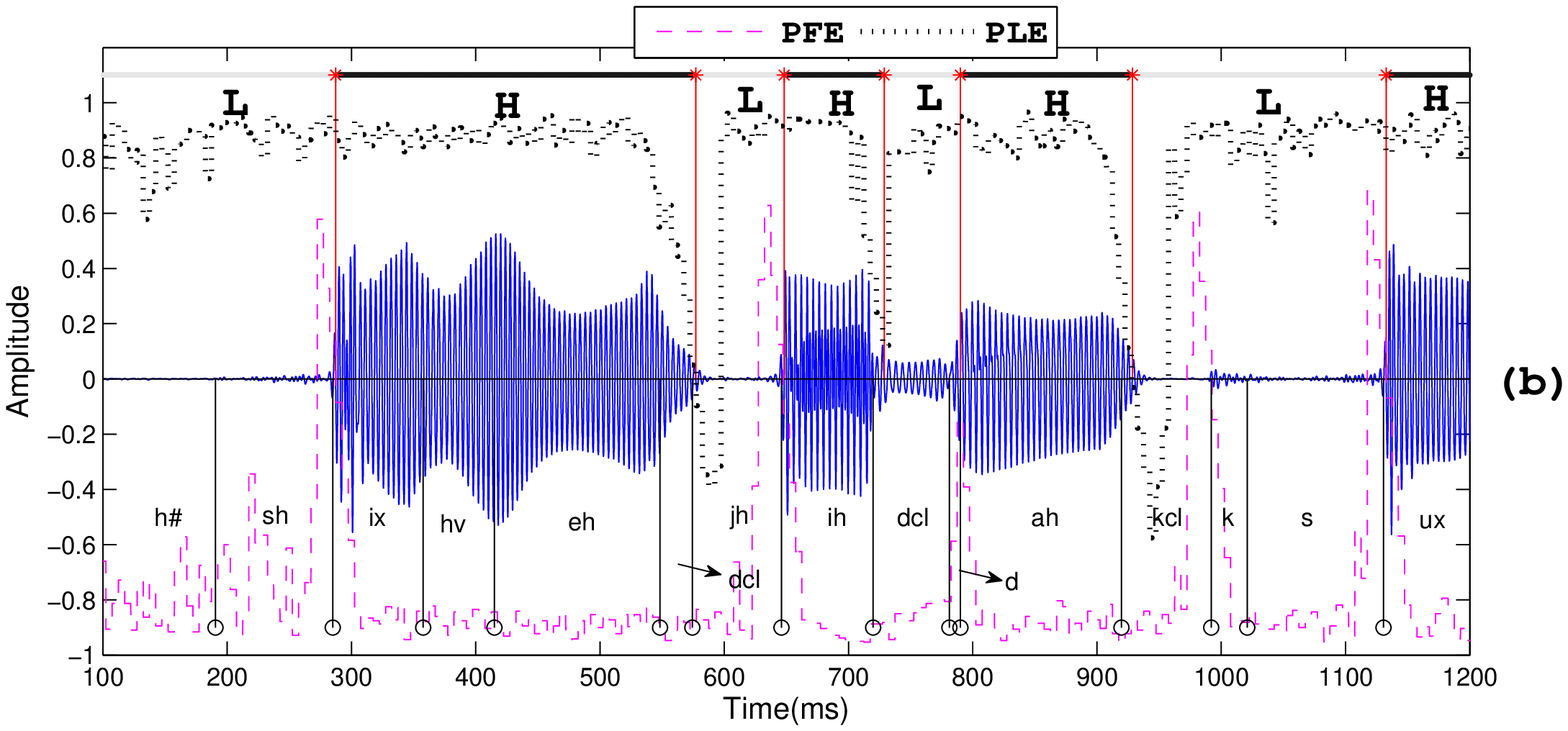}

\includegraphics[width=.75\textwidth,height=.22\textheight]{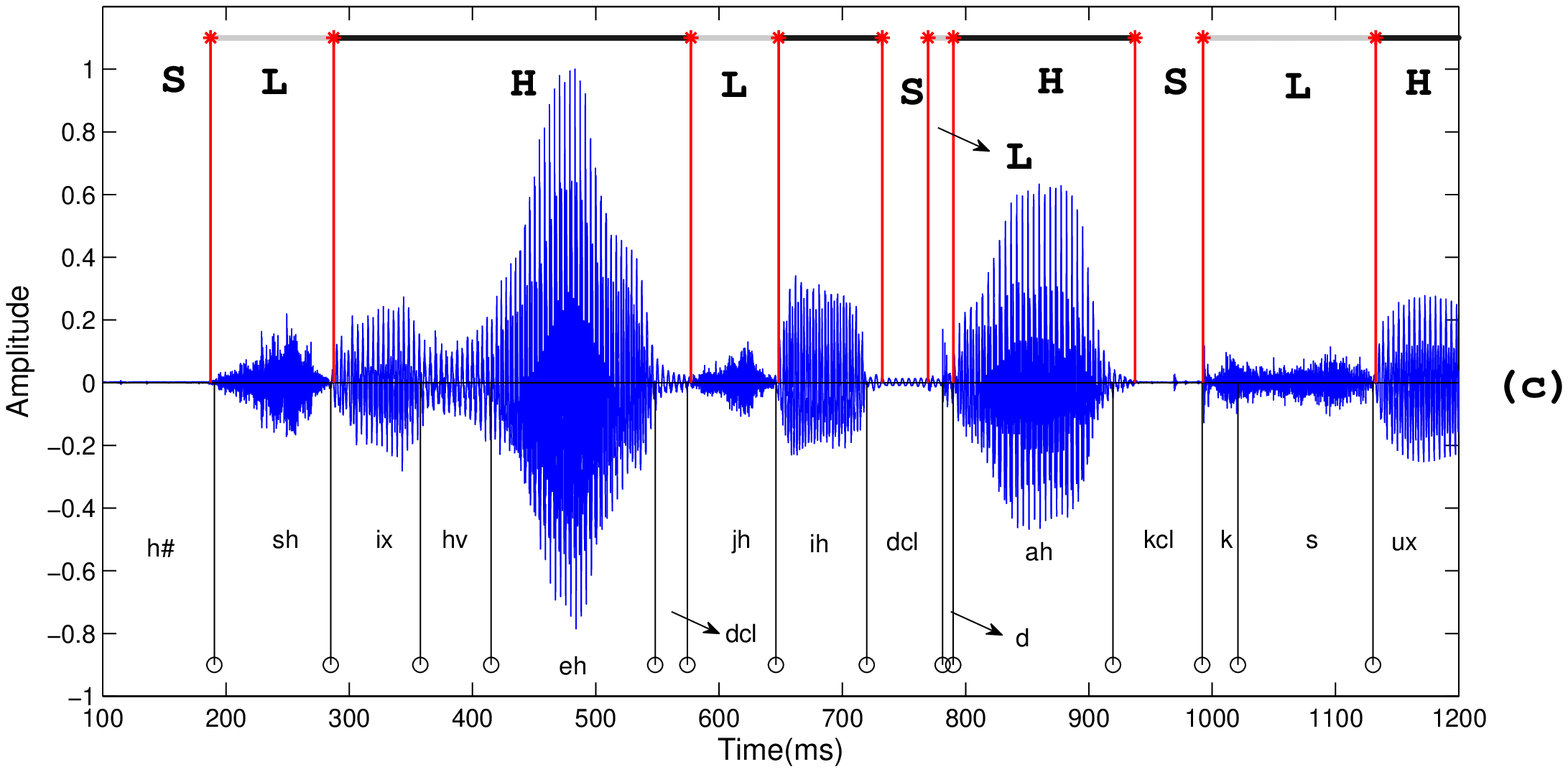}

\caption{Transitions (starred  markers) detected using (a) SI, (b) PFE/PLE  and (c) Merger Rule.}
\label{transs}
\end{figure*}

Fig.\ref{transs}(c) shows the class labels assigned after the application of merger rules. A S-N transition around 200 ms is followed by a L-H transition around 280 ms. Since the two transitions are spaced beyond 10 ms, both are retained. The N-class segment between S-N and L-H is assigned the L-class. The segment (approximately between 280 and 580 ms) consisting of /ix hv eh dcl/ belongs to a homogeneous H-class, since it lies between L-H and H-L transitions. Similarly, /ih/ (650 to 720 ms) and /ah/ (790 to 920 ms) belong to H-class. The segment /jh/ (580 to 650 ms) belongs to a homogeneous L-class. 

Now some situations are discussed, where merger rules are applicable. The boundary (around 730 ms) between /ih/ and /dcl/ is detected both as N-S and H-L transitions. However, since H-L transition was on the left of the N-S transition and within 20 ms (see Fig.\ref{transs}a), it is removed. Although the H-L transition is removed, the segment /ih/ between the L-H and N-S transitions is labeled as H-class. 

A S-N transition is detected across /dcl/ and /d/, and a L-H, across /d/ and /ah/. These two transitions, being 20 ms apart are retained as the temporal tolerance is 10 ms for a L-H following a S-N transition. Across /ah/ and /kcl/, both H-L and N-S are detected, in that order. However, since H-L transition is within 20 ms of N-S transition, the former is removed by the merger rule. Across /kcl-k/, only the S-N transition survives since L-H transition due to PFE is removed due to the low value of ADE. 

\section{Experimental details and evaluation }
 

The proposed AGR algorithm has been validated on the entire TIMIT database \cite{timit}, i.e., both training and test databases. It has been hand labeled at the phone level and the closure duration of stops have been explicitly marked. Accordingly, the class `stops' denotes `stop bursts'. It consists of several dialects of North American English, totaling 6300 utterances spoken by 630 speakers. 


Every detected transition is uniquely assigned to the nearest hand labeled boundary. The statistics of the temporal differences between the hand labeled boundaries and the assigned transition instants are computed. If the absolute temporal difference is less than a pre-defined temporal tolerance, then we say the boundary has been successfully detected.

In order to study the relationship between the manner of articulation and the homogeneous segments, the distribution of each class of phones among the five classes is computed.

For every sonorant and non-sonorant onset in the labelled database, we verify if there is a detected transition within a specified temporal tolerance. If no transition has been detected  within the tolerance for an onset, then it is a case of miss or deletion. This measures the accuracy of detection of onsets relative to the type of transition. A detected transition for which there is no associated hand labeled boundary is counted as an insertion. The ratio of the number of insertions to the total number of transitions detected is one of the performance measures.
 
\section{Experimental Results}

The results presented correspond to the total number of frames of about 3,818,197 and the total number of detected transitions of 144,715.

\subsection{Temporal accuracy of detection}

\begin{figure}
 
\includegraphics[width=.5\textwidth,height=.27\textheight]{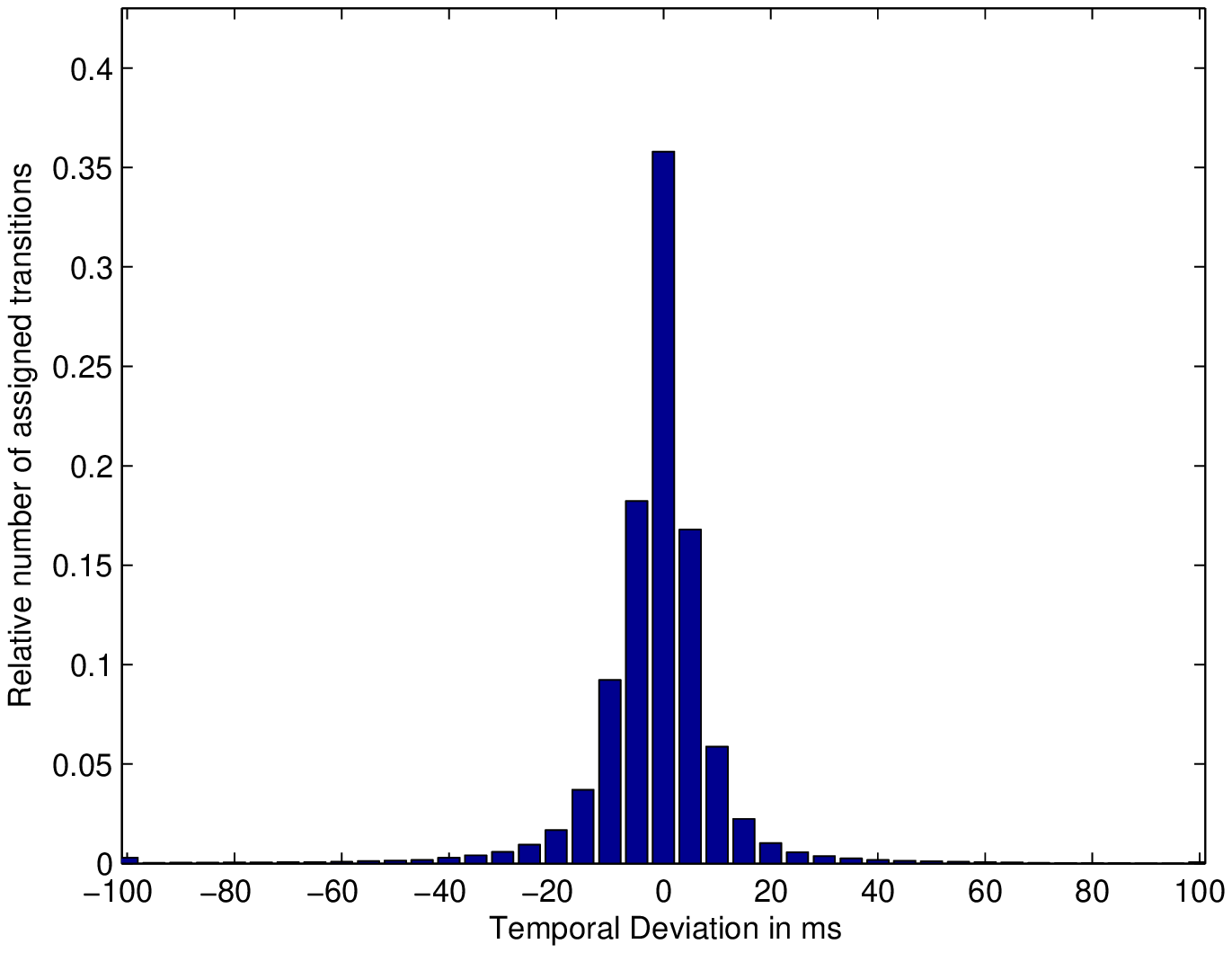}
 
\caption{Histogram of the temporal deviations of the detected transitions from the hand labeled boundaries.}
 \label{histacc}
 \end{figure}

Figure \ref{histacc} shows the histogram of the temporal deviations of the detected transitions from the hand labeled boundaries, using a bin size of 5 ms. The mean and standard deviation are -1.62 and 17.05 ms, respectively. We observe that 36.4\% of the detections are within $\underline{+}$2.5 ms. 

The detection accuracy is computed for different choices of the temporal tolerance, viz., 5 to 40 ms in steps of 5 ms. The ratio of successful detections to the total number of transitions, excluding insertions, is computed as the detection accuracy of transitions and is shown in Table \ref{tempacc} as a function of temporal tolerance. 57.8\% of the transitions lie within $\underline{+}$ 5 ms and 98\% of the transitions lie within $\underline{+}$ 40 ms. Thus, the temporal resolution of detection is higher than those of related previous works to be presented in Sec.\ref{comp}.

\begin{table}
\caption{Detection accuracy in percentage (\%) of transitions as a function of temporal tolerance in ms.}
\label{tempacc}
\resizebox{8.5cm}{!}{
\begin{tabular}{|c|c|c|c|c|c|c|c|c|}
\hline
Tolerance &5 &10  &15 &20  &25 &30 &35  &40 \\
\hline

Accuracy  &57.8 &80.7 &89.9 &93.6 &95.5 &96.7 &97.4 &98.0\\
\hline
\end{tabular}}
\end{table}

\subsection{Classes of phones detected across each type of transition  }
It is of interest to know the distribution of various phones which belong to the five classes H, L, S, HL and LH. For each of the classes of phones (vowels, semivowels etc.), the relative distribution of the duration of the phone classes amongst the five labelled classes is computed and is listed in Table \ref{reldistphn}. A temporal tolerance of 20 ms is assumed.

More than 91\% of vowels belong to H-class. But we note that there are about 4.8\% of vowels in L class and 0.5\% in S class. About 41\% of the `ax-h' phones lie in L-class, since it has the  characteristics of unvoiced speech, with a very low amplitude in the BPF signal. Amongst the semi-vowels, 74.0\% of `hh' lie in the L-class, since this phone also has characteristics similar to unvoiced speech.

It has been seen that in a nasal-fricative segment, there is a short interval of silence at the end of a nasal, which is however not hand labeled as silence. This explains the occurrence of about 7.2\% of nasals in S-class. A short silence segment is not unexpected since there is a change of source process as well as a drastic shift in the articulatory positions.

About 91\% of unvoiced fricatives and affricates lie in L-class. Affricates include the voiced affricate /jh/, which also lies in L-class. Amongst the fricatives, `th' behaves anomalously in that 20\% lies in S-class, since it sometimes manifests as a burst with a closure interval. 

56.9\% of voiced fricatives lie in L-class, whereas 28.6\% lie in H-class and 8.3\% go to S-class. The presence of voicing in voiced stops gives rise to a large amplitude BPF signal and when these classes follow a silence or a L-class phone, they go to H-class. 72\% of /z/ lies in L-class despite being a voiced fricative. The phone `dh' sometimes behaves like a stop with a closure and /v/ is realized both as voiced and unvoiced.

The phone labels of the TIMIT database are mapped to the phonetic classes, sonorants and non-sonorants. All vowels, semi-vowels and nasals are assigned to sonorant phonetic class. Non-sonorants include all the phones except the sonorants, others (`h\#', `epi', etc.) and the closures of stops. The relative distribution as per the phonetic classes, `sonorant', `non-sonorant' and `silence' is shown in Table \ref{reldistbrdcls}. 
About 89.8\% of sonorants lie in H-class. About 75.5\% of non-sonorants are in L-class. If we remove voiced fricatives and voiced stops from non-sonorants, then the unvoiced non-sonorants in L-class increases to 84\%. This suggests that we need two groups of non-sonorants. 84.2\% of `silence' segments lie in S-class with 10.5\% in L-class. Once again, this may arise due to some so called silence phones like `h\#' and `epi' having a high amplitude.

Based on the above results, we can broadly state that H-class represents the sonorant class and L-class represents unvoiced non-sonorants, whereas voiced non-sononrants may be found in both H and L classes.

\begin{table}
\centering
\caption{Relative distribution of each class of phones among the five classes. Results on the entire TIMIT data, containing both training and test data.}
\label{reldistphn}
\begin{tabular}{|c|c|c|c|c|c|}
\hline
Segment type &H &L &S &HL &LH\\
\hline
Vowels &91.8 &4.8 &0.5 &2.2 &0.7\\
\hline
Semivowels &86.2 &9.2 &1.3 &2.5 &0.9\\
\hline
Nasals &82.0 &7.2 &6.8 &3.3 &0.8\\
\hline
Unvoiced fricatives &4.2 &91.1 &2.0 &2.2 &0.4\\
\hline
Voiced fricatives &28.6 &56.9 &8.3 &5.0 &1.2\\
\hline
Voiced stops &48.0 &37.3 &12.1 &1.6 &1.1\\
\hline
Unvoiced stops &16.3 &70.5 &11.3 &1.3 &0.5\\
\hline
Affricates  &6.3 &91.0 &0.1 &1.8 &0.7\\
\hline
Others &4.5 &13.3 &82.1 &0.1 &0.0\\
\hline
Voiced closures &10.6 &8.7 &78.4 &1.7 &0.7\\
\hline
Unvoiced closures &2.1 &5.6 &92.0 &0.2 &0.1\\
\hline
\end{tabular}
\end{table}

\begin{table}

\caption{Distribution of each broad class of phones in the TIMIT database among the five classes}
\label{reldistbrdcls}
\centering
\begin{tabular}{|c|c|c|c|c|c|}
\hline
Phone class &H &L &S &HL &LH\\
\hline
Sonorant &89.8 &5.8 &1.3 &2.4 &0.7\\
\hline
Non-sonorant &15.2 &75.5 &6.2 &2.4 &0.7\\
\hline
Silence &4.8 &10.5 &84.2 &0.4 &0.1\\
\hline
Voiced non-sonorant &34.8 &50.6 &9.5 &3.9 &1.2\\
\hline
Unvoiced non-sonorant &8.4 &84.2 &5.0 &1.9 &0.5\\
\hline
\end{tabular}

\end{table}

   \subsection{Onset of sonorants and non-sonorants vis-a-vis the type of transition}       
   
   The onsets of sonorants and non-sonorants are considered as landmarks \cite{stevens}$^,$\cite{saloman}$^,$\cite{liu}. It would be of interest to relate the onsets of sonorants and non-sonorants to the detected types of transition. We have excluded /q/ from non-sonorants as done in several previous works\cite{apr}$^,$\cite{saloman}. Further, within the non-sonorants, we consider the fricatives and stop bursts separately to detect the onsets. The results are shown in Table \ref{pertransgrnd1}. For a tolerance of 30 ms, 94\% of onsets of sonorants occur at L-H or S-H transitions. We have considered sonorants following unvoiced fricatives, unvoiced stops and affricates, since voiced fricatives and voiced stops may lie in H class (see Table \ref{reldistphn}). The onsets of unvoiced fricatives and affricates occur at H-L and S-L transitions 85.4\% of the time within 30 ms. Stop closures are detected as onsets 80\% of the time across L-S and H-S transitions. Onsets of stop bursts invariably (88.7\%) follow a detected silence segment (S-H, S-L). The results are comparable even for a tolerance of 20 ms. Hence the proposed method also serves the purpose of landmark detection with a good accuracy and temporal resolution.

   \begin{table}

   \caption{Percentage of onsets of broad phonetic classes detected within a tolerance 20, 30 and 40 ms.}
   \label{pertransgrnd1}
      \resizebox{8.7cm}{!}{
\begin{tabular}{|c|c|c|c|c|}
\hline
Onset of &Type &20 &30 &40 \\
\hline
Sonorants+ &L-H, S-H &92.0 &94.0 &94.7\\
\hline
Unvoiced fricatives/affricates*& H-L, S-L  &83.0 &85.4 &86.5\\
\hline
Stop closures& L-S, H-S &77.2 &80.0 &81.4\\
\hline
Bursts &S-H, S-L &87.7 &88.7 &89.1\\
\hline
\end{tabular}
      }
      
      +Following an unvoiced fricative, unvoiced stop or an affricate.
       *Following a sonorant or a silence
      \end{table}

\subsection{Insertions}
The insertions on the whole TIMIT database are 8.7\%. About a third of these insertions occur during the silence, i.e, `others' and closure of stops. Segments like `h\#' and `epi' may contain impulse like noise with significant amplitude resulting in some spurious S-N and N-S transitions. About 24\% of the insertions occur during stops. These arise partly due to multiple bursts. A transition is also detected across a low level aspiration interval following a strong burst. While this is a desirable feature of the algorithm, since the aspiration interval is not explicitly marked, such transitions get reported as insertions. During unvoiced fricatives, especially, /f/, the amplitude of the signal varies considerably with intermittent low frequency, large amplitude pulses resulting in a high rate of insertions (about 11\%).
%

\subsection{\label{comp}Comparison with the previous work}

In terms of detecting classes, this work is comparable to manner classification\cite{juneja}$^,$\cite{hopkins} and in terms of detecting onsets, this work is closest to the landmark detection reported in the literature\cite{saloman}$^,$\cite{liu}. 

The present work differs from the previous related works in four important respects: (a) The temporal features used in this study are different from those proposed in the earlier studies. (b) The proposed algorithm has been tested on the entire TIMIT database, whereas the previous studies have reported results based on a limited test data (about 2000 tokens from the development set and 1200 from the test set of the TIMIT database in a study by Liu \cite{liu}; 504 utterances from the test set of the TIMIT database in Salomon et al\cite{saloman} study). (c) The transitions or landmarks to be detected correspond to different events. Liu\cite{liu} defined four landmarks and Salomon et al\cite{saloman} defined three landmarks. (d) The quoted results correspond to a temporal tolerance of 30 ms\cite{liu} or 50 ms\cite{saloman}.

Due to the above disparities, we can only make a broad qualitative comparison with the previous works. 

Salomon et al\cite{saloman} tested their method on the manner classes of sonorant, fricative, stop and silence. The average accuracy using 39 parameter MFCCs or 12 parameters of four temporal features is reported as 70\% for a tolerance of 50 ms, whereas with the combined features, it increases to 74.8\%. Compared to these results, the accuracies of the proposed method within 20 ms tolerance, are 89.8\%, 84.2\% and 84.2\% for sonorants, unvoiced non-sonorants and silence classes, respectively, when tested on the entire TIMIT database (see Table \ref{reldistbrdcls}). 

In Liu's\cite{liu} study, of the total number of landmarks, 83\% and 88\% were within 20 and 30 ms of the labeled boundaries, respectively. The classes considered in that study are sonorants, fricatives and bursts. The above results may be compared with the onset detection accuracies of the present work  (Table \ref{pertransgrnd1}). For a temporal tolerance of 30 ms, the onset accuracies for sonorants, unvoiced non-sonorants, closures and stop bursts are 94\%, 85\%, 80\% and 89\%  respectively. These results are as good or somewhat better than the above, though the types of landmarks and detected transitions are not strictly comparable to each other. 


\section{Conclusion}

For the DFs and PFs to be complimentary to the statistical approach, we believe that an acoustic-phonetics knowledge based approach has to be pursued. In our understanding, the highlights of such an approach are (i) It does not require a huge amount of training data. A small development set is considered sufficient. (ii) It makes use of acoustic correlates which are specific to the DF or PF being extracted, along with a simple rule based logic. In this paper, we have proposed a knowledge-based approach to the problem of detecting transitions in a speech signal. Further, several studies have pointed out the robustness of temporal features in speech perception\cite{saloman}$^,$\cite{rosen}. In the proposed method, using only four simple measures, we have been able to demonstrate that landmarks like the onsets of sonorants (L-H, S-H), unvoiced sonorants (H-L, S-L), closures of stops and stop bursts can be detected with a high accuracy ($>85$\%) and with a good temporal resolution (20 ms). These results are as good or better than state-of-the-art methods which make use of high dimensional acoustic features and sophisticated classifiers. Although a number of techniques exist for segmentation, alternate approaches are to be explored, since they may complement one another and offer robustness.

\subsection{Future Work}
During the course of this investigation, we have made some observations which are noted here for future work: (a) We could inquire how PFE/PLE measures perform instead of the abrupt energy change measures used in the literature for the detection of landmarks\cite{liu}, manner classes\cite{saloman}, bursts\cite{niyogi} and vowel onset points\cite{prasanna9}. (b) Our preliminary investigation shows that PFE and PLE measures computed on a speech signal, instead of bandpass signal, are useful to identify certain transitions within vocalic segments. Also, PFE and PLE may be computed on subband signals. (c) The number of extrema in a speech signal relative to the number of extrema in the corresponding bandpass signal is a useful parameter for distinguishing between voiced and unvoiced segments. (d) We have observed that bursts most often lie at the end of a silence or L-class. This narrows down the search interval for detecting the bursts. These preliminary observations have to be formalized and tested in a future work.

\end{document}